\documentclass{llncs}

\usepackage{geometry}
\usepackage{cite}
\usepackage{amsmath,amssymb,amsfonts}
\usepackage{algorithm}
\usepackage{algorithmicx} 
\usepackage{algpseudocode}
\usepackage{subfigure}
\PassOptionsToPackage{subfigure}{tocloft}
\usepackage{fixmetodonotes}
\usepackage{graphicx}
\usepackage{subfig}
\usepackage{textcomp}
\usepackage{xcolor}
\usepackage{bbm,bm}
\def\BibTeX{{\rm B\kern-.05em{\sc i\kern-.025em b}\kern-.08em
    T\kern-.1667em\lower.7ex\hbox{E}\kern-.125emX}}
    
\usepackage{subfigure,times,graphics,mathptm,epsfig,amsmath,xspace,endnotes,pifont,multirow,rotating,listings,amssymb,color,caption,nicefrac,adjustbox,todonotes,tabularx,mathtools,cite}

\author{Anousheh Gholami \inst{1} \and Nariman Torkzaban \inst{1} \and John S. Baras \inst{1} \and Chrysa Papagianni \inst{2}\\
\institute{ Department of Electrical and Computer Engineering and the Institute for Systems Research\\
University of Maryland, College Park, MD 20742, US\\
\email{\{anousheh | narimant | baras\} @umd.edu}  \and
Nokia Bell Labs, Antwerp, Belgium\\ \email{chrysa.papagianni@nokia-bell-labs.com}}
}

\begin{document}
%
\title{Joint Mobility-Aware UAV Placement and Routing in Multi-Hop UAV Relaying Systems}

\maketitle

\begin{abstract}
Unmanned Aerial Vehicles (UAVs) have been extensively utilized to provide wireless connectivity in rural and under-developed areas, enhance network capacity and provide support for peaks or unexpected surges in user demand, mainly due to their fast deployment, cost-efficiency and superior communication performance resulting from Line of Sight (LoS)-dominated wireless channels. In order to exploit the benefits of UAVs as base stations or relays in a mobile network, a major challenge is to determine the optimal UAV placement and relocation strategy with respect to the mobility and traffic patterns of the ground network nodes. Moreover, considering that the UAVs form a multi-hop aerial network, capacity and connectivity constraints have significant impacts on the end-to-end network performance. To this end, we formulate the joint UAV placement and routing problem as a Mixed Integer Linear Program (MILP) and propose an approximation that leads to a LP rounding algorithm and achieves a balance between time-complexity and optimality.


\end{abstract}

\section{Introduction}
Over the past decade, UAVs have been adopted in a broad range of application domains, due to their autonomy, high mobility and low cost. Historically, UAVs have been primarily used in the military, usually deployed in hostile territory to reduce risk for aircrew. Recent advances in UAV technologies have made them more affordable and accessible to civilian and commercial applications such as cargo transport, emergency search and rescue, precision agriculture, commercial package deliveries, etc. Moreover, UAVs are seen as a promising solution for next generation wireless networks because of their inherent advantages, including flexible and fast deployment and reconfiguration, as well as a higher chance of having Line-of-Sight (LoS) links leading to less impaired communication channels compared to terrestrial wireless communication systems. 
\begin{figure}[t]
\centering
\includegraphics[width=1\textwidth]{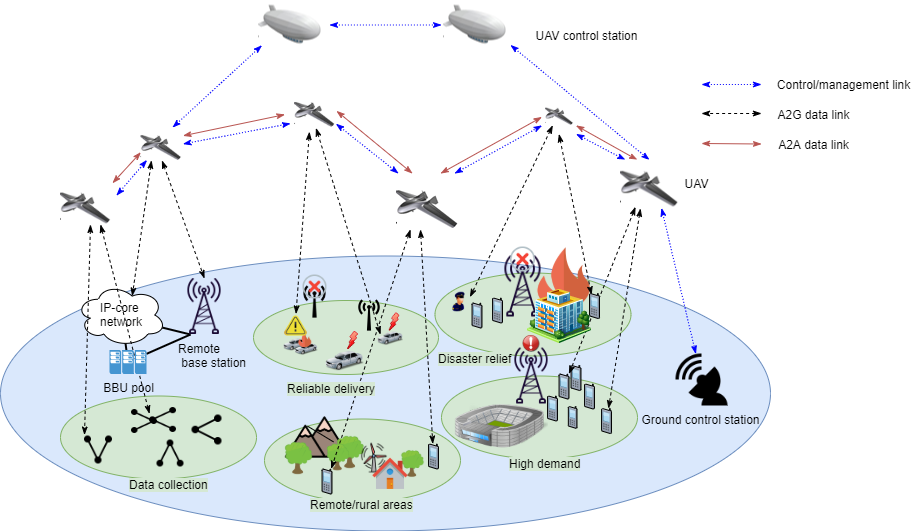}
  \caption{The architecture and application scenarios of a multi-UAV relay system}
    \label{fig:usecase}
\end{figure}

According to \cite{zeng2016wireless}, UAV-aided wireless communications can fall into three representative categories of use cases; (i) UAV-aided ubiquitous coverage, (ii) UAV-aided information dissemination and data collection and (iii) UAV-aided relaying. Focusing on the latter, communication relaying is an effective technique for network coverage extension and throughput maximization. However, a number of key challenges should be addressed in order to use UAVs as mobile relaying nodes, providing broadband communication to users (or user groups) without direct and/or reliable communication links (e.g., in disaster-hit or rural areas \cite{gholami2019drone}). First, efficient algorithms should be devised to place UAVs in a 3D space. The mobility of the UAVs introduces new challenges to the network design problem compared to the traditional static relaying and fixed-infrastructure schemes (e.g. WiFi access-points). Moreover, in order to cover large geographical areas and because of the limited transmission range of UAVs, a swarm of UAVs is needed to route users' traffic demands through wireless multi-hop path(s). Due to the intermittent wireless links and frequent topology changes in such mobile ad hoc networks (MANETs), the traffic routing decision should be considered together with the UAVs placement and relocation. More importantly, such decisions should be adaptive to the topology and traffic pattern changes in a timely manner.

In this study, we consider an aerial platform consisting of multiple UAVs that supports the traffic demand of a ground network. Multi-hop relaying in the next generation of wireless networks will not only facilitate the coverage of more UEs and the support of long-distance communications, but also will be able to handle overloaded networks. Fig. \ref{fig:usecase} illustrates the architecture and application scenarios of such systems. In contrast to the majority of existing studies such as \cite{zhang2017joint} and \cite{zeng2016throughput} that solely focus on a single UAV relay and the air-to-ground (A2G) access links with one-hop communication or at most two-hop communications considering also the UAV-to-BS links in UAV-aided cellular networks (e.g. \cite{lyu2018uav}), we exploit a multi-hop aerial relaying platform. Moreover, the rate-constrained UAV-to-UAV (or Air-to-Air, A2A) communications and the connectivity between UAVs are considered in the proposed framework. Although facility placement and traffic routing are usually addressed sequentially as two separate problems, the meaningful interrelation between the two, as discussed in \cite{9149175} makes it more reasonable to approach them in a single model. Thus, another contribution of this manuscript is jointly optimizing the UAV placement and the routing decisions, when the ground network is quasi-static. We also extend our approach to the case of a mobile ground network and consider the impact of the UAVs' speed, while most of the existing relaying schemes utilize static relay nodes due to the practical constraints on relay mobility and the need for high-throughput links \cite{zeng2016throughput}. In order to reduce the system energy consumption and since the propulsion energy consumption of UAVs is typically significantly greater than the energy consumption for communications \cite{8660516}, we avoid unnecessary UAVs' relocation in subsequent snapshots in our solution. The problem is formulated as a mixed-integer linear program (MILP). To reduce the time-complexity and enable real time re-positioning of the mobile relays and routing decision, we propose an approximation algorithm using linear programming (LP) relaxation and a rounding procedure. The proposed approach assumes logically centralized network control i.e., software defined networking (SDN). The controller's global network view in an SDN architecture, renders centralized UAVs placement and adaptive routing strategies feasible \cite{li2018uav}. The controller may be placed at a remote ground center, inside the ad hoc network devices as in \cite{poularakis2019hybrid} or at UAVs as shown in Fig \ref{fig:usecase}.

The paper is organized as follows. Section \ref{sec:sys} describes  the  problem and the system model. In section \ref{sec:formulation}, we introduce a MILP formulation for the optimal UAV placement in the case of a quasi-static ground network and then extend our approach, considering mobility of the nodes, as well as provide our LP-based approximation method. Performance evaluation is presented in section \ref{sec:evaluation}, while we provide the overview of the related work in section \ref{sec:relatedwork}. Finally, in section \ref{sec:conclusions}, we highlight our conclusions and discuss our future work.

\section{System Model and Problem Description}
\label{sec:sys}
\subsection{Radio Propagation Model}

We adopt the model proposed in \cite{al2014optimal} for the A2G propagation model, where two signal propagation groups are considered; Line-of-Sight (LoS) and Non-Line-of-Sight (NLoS). The latter corresponds to receivers with no Line-of-Sight but still having coverage via strong reflections and diffraction. Additional impairments to the radio channel are caused by scattering and shadowing from the man-made structures in the environment. The occurrence probability of LoS is given by:
\begin{align}
    p_{LoS} = \frac{1}{1+a  exp(-b(\frac{180}{\pi}tan^{-1}(\frac{h}{r_{n,l}})-a))}
\end{align}
where $a$ and $b$ are constants depending on the environment, $h$ is the UAV altitude and $r_{n,l}$ is the horizontal euclidean distance between the UAV $l$ and the user equipment (UE) $n$. 
The probability of NLoS is $p_{NLoS} = 1 - p_{LoS}$ and the total A2G path loss (in dB) as a function of $r_{n,l}$ and $h$ is:
\begin{align}
\label{loss power}
    L(h,r_{n,l}) = p_{LoS}L_{LoS} + p_{NLoS} L_{NLoS}, \\
    L_{LoS} = 20log(\frac{4\pi f_c d_{n,l}}{c})+\eta_{LoS} \nonumber, \\
    L_{NLoS} = 20log(\frac{4\pi f_c d_{n,l}}{c})+\eta_{NLoS} \nonumber
\end{align}
where $f_c $ is the carrier frequency, $d_{n,l} = \sqrt{h^2 + r_{n,l}^2}$ is the distance between UAV $l$ and UE $n$, $\eta_{LoS}$ and $\eta_{NLoS}$ are respectively the average additional losses due to the environment. 

We assume that the A2A links are dominated by LoS components resulting in the following free space path loss model: 
\begin{align}
    \label{uav-uav loss}
    L(r_{u,v}) = 20log(\frac{4\pi f_c r_{u,v}}{c})
\end{align}
where $r_{u,v}$ is the distance between UAV $u$ and UAV $v$. Assuming that an interference-coordination mechanism among adjacent UAVs and users is available, the interference is negligible and the received signal at a node is only affected by Additive White Gaussian Noise (AWGN). Consequently, the coverage radii for the A2G and A2A channels, denoted by $R_1$ and $R_2$, satisfy:
\begin{align}
    P_{UE}= L(h,R_1)+\gamma_{min} + \sigma_n^2
    \label{R1}
\end{align}
\begin{align}
    P_{UAV}=  20log(\frac{4\pi f_c R_2}{c}) + \gamma_{min} + \sigma_n^2
    \label{R2}
\end{align}
The QoS requirement is expressed in terms of the minimum received SNR at the receiver ($\gamma_{min}$), noise power ($\sigma_n$) and maximum transmission power of UAVs ($P_{UAV}$) and users ($P_{UE}$), where $P_{UE} \leq P_{UAV}$.

\subsection{Problem Description}
We discretize time and consider a directed graph $\mathcal{G}^t = (\mathcal{V}^t,\mathcal{E}^t)$  representing the topology of the ground network at snapshot $t$. 
The vertex set $\mathcal{V}^t$ represents the network nodes and the edge set $\mathcal{E}^t$ represents the wireless links, i.e. $(u,v) \in \mathcal{E}^t$ if and only if node $v$ can receive data packets directly from node $u$. We assume that all node-to-node communication is unicast, i.e. each packet transmitted by a node $u \in \mathcal{V}^t$ is intended for a unique $v \in \mathcal{V}^t$ where $(u,v) \in \mathcal{E}^t$. Moreover, each wireless link has a maximum capacity $c_{uv}$. For the sake of simplicity, the superscript $t$ is dropped in the following.

There are traffic demands between UEs given by a  traffic demand matrix $D$, where the element $D_{uv}$ denotes the amount of demand from the source UE $u$ to the destination UE $v$. The demand profile can be estimated using existing MANET traffic pattern inference (if a central controller is not available) or the schemes proposed for cellular networks (e.g, \cite{zhang2019deep}, \cite{yu2015modeling}) which are also applicable here in the presence of the UAV or ground control stations. Due to the limited transmission power of the UEs and in order to reduce the control traffic overhead required for traffic routing, the ground network is partitioned into $M$ clusters, denoted by $C_i$, $i = 1,...,M$. Each cluster has a cluster head (CH) which functions as a  gateway, relaying the cluster total traffic to the aerial platform. Let $CH_i$, $i = 1,...,M$ denote the $i$th cluster head.  Given $D$, the inter-cluster traffic demand for a pair of clusters and from $CH_i$ to $CH_j$ is calculated by: 
$$TD_{ij} = \sum_{u\in C_i, v\in C_j} D_{uv}, \quad \forall i , j \in \{CH_1,...,CH_M\}$$
We denote by $i\rightarrow{}j$, a traffic flow originated from $CH_i$ and destined to $CH_j$. 
The problem considered in this paper entails the optimal placement of at most $N_{max}$ available UAVs as relay nodes to support the traffic demand of the ground origin destination (OD) flows. For each OD flow $i\rightarrow{}j$, a collection of aerial multi-hop paths can be used to route the traffic demand of the flow.

Let $\mathcal{U} = \{u_i, i = 1,..., |\mathcal{U}|\}$ denote the set of potential locations for UAV placement, where $v_i$ stands for the $i$th location. Here, we assume that all UAVs are placed at the same altitude $h$; however, it is easy to extend the formulation to a 3D UAV placement where the set $\mathcal{U}$ includes locations at different heights. The following graphs are defined for the problem formulation:\\\\
\textbf{Demand Graph}: We model the connectivity and traffic requirements of the ground clusters by a directed graph $\mathcal{G}_D = (\mathcal{V}_D, \mathcal{E}_D)$ where, $\mathcal{V}_D=\{CH_1,...,CH_M\}$ is the set of all CHs, and $(i,j) \in \mathcal{E}_D $ if and only if the OD flow $i\rightarrow{}j$ exists for $i, j \in \mathcal{V}_D$.
\\\\\textbf{Network Graph}: We introduce a directed graph $\mathcal{G}_P = (\mathcal{V}_P,\mathcal{E}_P)$ where $\mathcal{V}_P = \mathcal{V}_{D}\cup \mathcal{U}$ and $(u,v)\in \mathcal{E}_P$ if and only if $d_{u,v}\leq R_2$ for A2A links, and $d_{u,v}\leq R_1$ for A2G links.

\section{Problem formulation}
In this section, the MILP formulations for the optimal UAV placement and traffic routing in both static and mobile ground networks are presented and the proposed LP-based approximation solution is discussed.
\label{sec:formulation}
\subsection{MILP formulation}
Given the network and demand graphs $\mathcal{G}_P, \mathcal{G}_D$, we formulate the problem at hand considering the following decision variables: 
 \begin{itemize}
 \item A set of binary variables $\bm{x}$, where $x_u$ is set to 1 if a UAV is deployed at position $u \in \mathcal{U}$ and $0$ otherwise.
 \item A set of continuous variables $\bm{f}$, where $f_{uv}^{ij}$ is the amount of traffic from OD flow $i\xrightarrow{}j$ assigned to the link $(u,v) \in \mathcal{E}_P$.
 \item A set of continuous variables $\bm{y}$, where $y_{ij}$ denotes the traffic amount of the OD flow $i\xrightarrow{}j$ that is not supported (not delivered).
 \end{itemize}
A summary of the system model parameters and variables is given in Table \ref{param-var}. The proposed MILP formulation for the joint UAV placement and traffic routing (\textbf{UPR}\_\textbf{MILP}$_{}$) is as follows:
\begin{table}[h]
\centering
\caption{System model parameters and variables}
\begin{center}
\scalebox{0.8}{
\begin{tabular}{|c|c|}
\hline
\hline
Variables & Description\\
\hline
$x_u$ & Binary decision variable of UAV placement at position $u$\\
$f_{ij}^{uv}$ & The amount of $(i,j)$ traffic d assigned to $(u,v)$\\
$y_{ij}$ & Total unsupported traffic of the OD pair $(i,j)$\\
\hline
Parameters & Description\\
\hline
$\mathcal{G}_P = (\mathcal{V}_P, \mathcal{E}_P)$ & The network graph of UAVs and ground cluster heads\\
$\mathcal{G}_D = (\mathcal{V}_D, \mathcal{E}_D)$ & The demand graph\\
$M$ & Number of ground cluster heads\\
$\mathcal{U}$ & The set of UAV potential locations\\
$N_{max}$ & Available number of UAVs\\
$h$ & UAVs height\\
$D$ & Traffic demand matrix of the ground network\\
$TD_{ij}$ & Traffic demand between the CHs $i$ and $j$\\
$c_{uv}$ & Capacity of the link $(u,v)$\\
\hline
\hline
\end{tabular}}
\end{center}
\label{param-var}
\end{table}
\begin{align}
&\textbf{minimize}
\quad \phi\frac{\sum_{u\in\mathcal{U}} x_u}{N_{max}} + (1-\phi) \frac{\sum_{(i,j)\in \mathcal{E}_D} y_{ij}}{\sum_{(i,j)\in\mathcal{E}_D}TD_{ij}} 
\label{obj} \\
&\textbf{Feasibility Constraints:} \nonumber \\
& f^{ij}_{uv} \leq x_u c_{uv} \quad \forall (i,j) \in \mathcal{E}_D, u\in \mathcal{U}, v\in \mathcal{V}_{p} \label{feas1} \\
& f^{ij}_{uv} \leq x_v c_{uv}\quad \forall (i,j) \in \mathcal{E}_D, v \in \mathcal{U}, u\in \mathcal{V}_{p} \label{feas2}\\
&\sum_{u\in\mathcal{U}} x_u \leq N_{max} \label{feas3}\\
&\textbf{Flow Constraints:} \nonumber \\ 
&\sum_{v\in \mathcal{V}_{P}}^{} (f^{ij}_{uv} - f^{ij}_{vu}) =   \left\{
  \begin{array}{@{}ll@{}}
    0 & \forall u \in \mathcal{V}_{P}  \symbol{92} \{i,j\}, (i,j) \in \mathcal{V}_D\\
    TD_{ij}-y_{ij} & u = i, \forall(i,j) \in \mathcal{E}_D\\
    -(TD_{ij}-y_{ij}) & u = j, \forall(i,j) \in \mathcal{E}_D\\
  \end{array}\right.\label{flow}\\
&\textbf{Capacity Constraints:} \nonumber \\
&\sum_{(i,j)\in \mathcal{E}_D} f_{uv}^{ij} \leq c_{uv} \quad \forall (u,v) \in \mathcal{E}_P \label{cap}\\
&\textbf{Domain Constrains:} \nonumber\\
& 0 \leq f^{ij}_{uv} \quad \forall i,j \in \mathcal{V}_D, u,v \in \mathcal{V}_p \label{dom1}\\
& 0 \leq y_{ij} \quad \forall (i,j) \in \mathcal{E}_D \label{dom2}\\
& x_u \in \{0,1\} \quad \forall u \in \mathcal{U} \label{dom3}
\end{align}
The objective function \eqref{obj} aims at jointly minimizing the cost of UAV deployment (reflected as the number of deployed UAVs) and the total amount of requested traffic that can not be supported by the network (the total unsupported traffic). We normalize both metrics to be between $0$ and $1$ in order to avoid the known problem of different range values in Pareto Analysis (i.e. one metric having large value and the other one having small value). Since we have in our formulation two performance objectives (minimizing the number of deployed UAVs and minimizing the unsupported traffic), a full solution of the problem requires the complete tradeoff analysis between these two metrics and finding the Pareto Points or Pareto Frontier of this tradeoff problem. To arrive at equation \eqref{obj}, we employed what is known as the ``scalarization method'' for tradeoff analysis. This method is less computationally intensive. To fully understand the tradeoff between these two metrics using the scalarization method, we need to vary $\phi$ between 0 and 1. In this way we can compute the convexified Pareto Frontier. Indeed in our experiments, we tested different values for $\phi$ and selected a relatively small value to promote a solution that primarily enhances the performance of the network by minimizing the unsupported traffic.

Constraints \eqref{feas1} and \eqref{feas2} guarantee that the amount of traffic assigned to an A2G link is nonzero only if an UAV is placed at the aerial end of the link. Constraint \eqref{feas3} limits the maximum number of deployed UAVs, while constraints \eqref{flow} enforce flow conservation, i.e. the sum of all inbound and outbound traffic for the UAV relays should be zero. Moreover this constraint ensures that for each OD flow $i\rightarrow j$, the inbound (outbound) traffic to $j$ (from $i$) is $TD_{ij}-y_{ij}$ (the amount of supported traffic). Constraints \eqref{cap} ensure that the total traffic assigned to a link does not exceed its capacity. Finally, \eqref{dom1}, \eqref{dom2} and \eqref{dom3} express the domain constraints.
\subsection{MILP formulation with UAV mobility constraints}
In the case of mobile UEs or dynamic traffic patterns, \textbf{UPR}\_\textbf{MILP} can be reapplied periodically in order to update the UAV positions \cite{perumal2008aerial}.  This update rate can be in the order of seconds \cite{han2009optimization}. To consider the effect of UAVs maximum speed in a dynamic environment, we add mobility constraints to the optimization problem discussed in the previous section. The maximum speed of UAVs is represented by $v_{max}$ and the time duration of a snapshot is denoted by $\Delta T$. For each $u_i\in\mathcal{U}$, let $\mathcal{B}_i \subset \mathcal{U}$ denote the set of potential locations that the UAV deployed in $u_i$ can reach in one snapshot, i.e. $u_j\in \mathcal{B}_i$ if $d_{i,j} \leq v_{max}\Delta T$. 

Given $\mathcal{B}_i$ and the UAV placement decision variables at snapshot $t-1$ ($\boldsymbol{x}^{t-1}$), the UAV mobility constraints at snapshot $t$ can be expressed as: 
\begin{equation}
    \sum_{u_j \in \mathcal{B}_i} x_j^{t} \geq \bbbone{\{x_i^{t-1}=1\}}
    \label{mob}
\end{equation}
Moreover, in order to reduce the propulsion energy consumption of UAVs by avoiding unnecessary and less-effective UAV relocations in consecutive snapshots, we add another term to the objective function \eqref{obj}. The new objective function is:
\begin{equation}
    (\phi\frac{\sum_{u\in\mathcal{U}} x_u}{N_{max}} + (1-\phi) \frac{\sum_{(i,j)\in \mathcal{E}_D} y_{ij}}{\sum_{(i,j)\in\mathcal{E}_D}TD_{ij}})+ \alpha (max_{u}|x_u^t-x_u^{t-1}|)
    \label{obj2}
\end{equation}
where $\alpha$ is a constant factor determining the balance between the two terms of the objective function. Instead of the maximum function in the new objective and in order to get rid of the absolute value, we define a scalar variable $z$ and add it to the objective function as follows:
\begin{equation}
    (\phi\frac{\sum_{u\in\mathcal{U}} x_u}{N_{max}} + (1-\phi) \frac{\sum_{(i,j)\in \mathcal{E}_D} y_{ij}}{\sum_{(i,j)\in\mathcal{E}_D}TD_{ij}})+ \alpha z
    \label{obj3}
\end{equation}
and we add the following set of constraints to the optimization problem:
\begin{equation}
    x_u^t - x_u^{t-1} \leq z \quad \forall u \in \mathcal{U}
    \label{movement1}
\end{equation}
\begin{equation}
    x_u^{t-1} - x_u^t \leq z \quad \forall u \in \mathcal{U}
    \label{movement2}
\end{equation}
The resulting MILP is referred to as \textbf{MUPR}\_\textbf{MILP} and is an NP-hard problem. However, the decision variables have to be determined in real-time, in response to the network changes. In the subsequent section we employ an LP-relaxation to deal with the time-complexity of the \textbf{MUPR}\_\textbf{MILP}. A greedy rounding approach is used to obtain the binary solution of the original problem.

\subsection{LP relaxation and rounding algorithm}
We derive the Linear Programming (LP) model of \textbf{MUPR}\_\textbf{MILP} by relaxing the binary variables $x_u^t$ or replacing the constraint sets \eqref{dom3} by:
\begin{equation}
x_u^t \in [0,1],  \forall u\in\mathcal{U}
\end{equation}
The resulting LP is represented by \textbf{UPR}\_\textbf{LP}. We also define the set $\mathcal{X} \subseteq \mathcal{U}$ based on which the following LP denoted by \textbf{UPR}\_\textbf{LP}\_\textbf{reduced}($\mathcal{X}$) is defined:
\begin{align}
    &\textbf{minimize} \quad (\phi\frac{\sum_{u\in\mathcal{U}} x_u^t}{N_{max}} + (1-\phi) \frac{\sum_{(i,j)\in \mathcal{E}_D} y_{ij}}{\sum_{(i,j)\in\mathcal{E}_D}TD_{ij}} ) + \alpha z
    \\
    &\text{s.t} \quad \eqref{feas1}-\eqref{dom1}, \eqref{movement1},\eqref{movement2}
    \\
    & x_u^t = 1, \quad \forall u \in \mathcal{X}
    \\
    & x_u^t \in [0,1] \quad \forall u \notin \mathcal{X}
\end{align}
We introduce a rounding-based decision-making process (DM-LP) to retrieve the binary decision variables of \textbf{MUPR}\_\textbf{MILP} at each snapshot by solving a sequence of \textbf{UPR}\_\textbf{LP}\_\textbf{reduced}($\mathcal{X}$) problems iteratively. Similar approach has been used for solving MILPs in resource allocation problems such as \cite{torkzaban2019trust}. The proposed solution is shown in \textbf{Algorithm\ref{det-algo}}. The set $\mathcal{X}$ represents the locations chosen for UAV placement and is updated within each iteration (line \eqref{algo:X}). The final $\mathcal{X}$ reflects the UAV placement decision. As explained in lines \eqref{algo:S}-\eqref{algo:n1}, UAVs are placed deterministically with the priority given to the neighboring locations of the deployed UAVs at the previous snapshot (reflected in the definition of the set $S$ which is constructed by the union of the sets $\mathcal{B}_v$ for $ v\in \mathcal{U}:x_{v}^{t-1}=1$) in order to not violate the mobility constraints. For example, if two UAVs are placed at locations $u_1, u_2$ at snapshot $t-1$, the set $\mathcal{B}_1\cup \mathcal{B}_2$ is first considered for UAV placement at snapshot $t$ so that at least one UAV is placed at one of the locations of $\mathcal{B}_1$ (similarly for $\mathcal{B}_2$). Once all mobility constraints are satisfied, all the remaining potential UAV locations are considered for the placement of new UAVs. In both cases, a UAV is deployed at the position with maximum $x$ value within each iteration (line \eqref{algo:n1} and \eqref{algo:n2}). The algorithm terminates when the addition of a new UAV does not reduce the objective function or makes the problem infeasible, i.e. $x_{u}^t = 0, \forall u \in \mathcal{X}$ or equivalently, $x_{u^*}^t = 0$. Finally, the routing decisions are automatically obtained from the $\boldsymbol{f}$ solution of the last iteration. Moreover, the input $\bm{x}$ is $\bm{0}$ in the first snapshot meaning that any location in $\mathcal{U}$ can initially be considered for UAV placement.
\begin{algorithm}
 \caption{DM-LP}
 \label{det-algo}
 \begin{algorithmic}[1]
 \Statex \textbf{Input: } $\mathcal{G}_p^t$, $\mathcal{G}_D^t$, $ TD^t$,  $\bm{x}^{t-1}$
 \Statex \textbf{Output: } $\bm{x}^t$, $\boldsymbol{f}$ 
 \State Initialize $\mathcal{X} \gets \emptyset$, $Terminate \gets False$ 
 \Repeat 
 \If{not first iteration} \State $\mathcal{X} \gets \mathcal{X} \cup \{u^*\}$ \label{algo:X} 
 \EndIf
 \State $\{x^t_u, f^{ij}_{uv}\}\xleftarrow{}$Solve \textbf{UPR}\_\textbf{LP}\_\textbf{reduced}($\mathcal{X}$) \label{algo:LP}   
 \State $S \gets \cup_{\{v:x_{v}^{t-1}=1, v\notin \mathcal{X}\}} \{argmax_{k\in\mathcal{B}_v}$ $x^t_{k}\}$ \label{algo:S} 
 \If{$S \neq \emptyset$ }\Comment{ Mobility constraints are not satisfied }
 \State $u^* =$ $argmax_{u\in S}$ $x_u^t$  \label{algo:n1} 
 \Else \Comment{ Mobility constraints are satisfied} 
 \State $u^* =$ $argmax_{u\notin\mathcal{X}}$ $x_u^t$  \label{algo:n2}  
 \EndIf
 \If{$x_{u^*}^t > 0$}  
 \State $x_{u^*}^t = 1$
 \Else 
 \State $Terminate \gets True$ 
 \EndIf 
 \Until $Terminate == False$\\
 \Return $\bm{x}^t$, $\boldsymbol{f}$
 \end{algorithmic} 
 \end{algorithm}

It is important to note that compared to \textbf{MUPR}\_\textbf{MILP} which is intractable for large networks, the proposed approximation algorithm calls the LP solver at most $|N_{max}|+1$ times. In the next section, we provide numerical results to evaluate the performance of the proposed approach.

\section{Performance Evaluation}\label{sec:evaluation}
\begin{figure*}[t]
\begin{center}
\begin{minipage}[h]{0.45\textwidth}
  \includegraphics[width=1\textwidth]{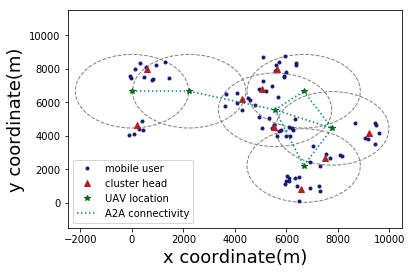}
  \caption{DM-MILP UAV placement}
    \label{fig:placement_milp}
\end{minipage}
\begin{minipage}[h]{0.45\textwidth}
  \includegraphics[width=1\textwidth]{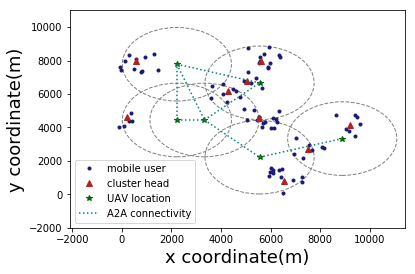}
  \caption{DM-LP UAV placement}
    \label{fig:placement_lp}
\end{minipage}
\begin{minipage}[h]{0.45\textwidth}
  \includegraphics[width=1\textwidth]{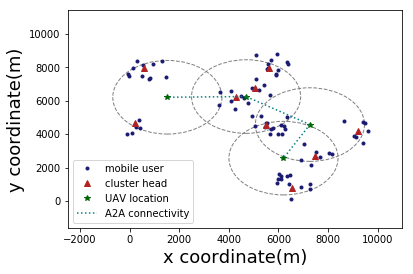}
  \caption{DM-Conn UAV placement \cite{perumal2008aerial}}
    \label{fig:placement_senni}
\end{minipage}
\end{center}
\end{figure*}

\begin{figure*}[t]
\begin{center}
\begin{minipage}[h]{0.4\textwidth}
  \includegraphics[width=1\textwidth]{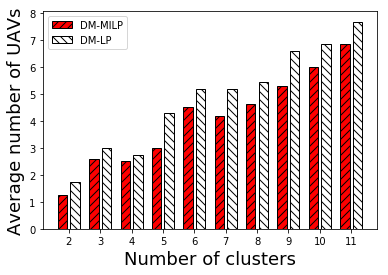}
  \caption{Number of UAVs: DM-LP vs. DM-MILP}
    \label{number}
\end{minipage}
\hspace{1em}
\begin{minipage}[h]{0.4\textwidth}
  \includegraphics[width=1\textwidth]{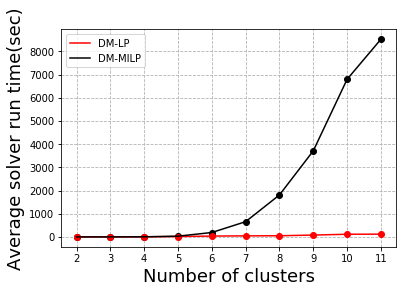}
  \caption{Solver runtime: DM-LP vs DM- MILP}
    \label{fig:time}
\end{minipage}
\hspace{1em}
\begin{minipage}[h]{0.4\textwidth}
  \includegraphics[width=1\textwidth]{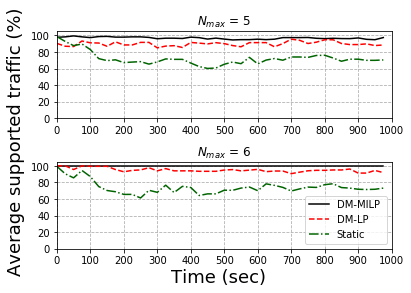}
  \caption{Average percentage of supported traffic}
    \label{fig:sup}
\end{minipage}
\hspace{1em}
\hspace{1em}
\begin{minipage}[h]{0.4\textwidth}
  \includegraphics[width=1\textwidth]{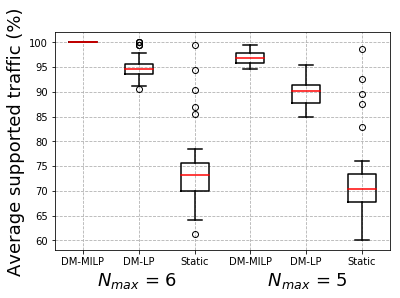}
  \caption{Average percentage of supported traffic profile}
    \label{fig:bp}
\end{minipage}
\end{center}
\end{figure*}

In this section, we benchmark out proposed decision-making process, DM-LP, against the exact solution, denoted as DM-MILP and the connectivity-based approach proposed in \cite{perumal2008aerial}, namely DM-Conn. We also compare the performance of the mobile and static UAV deployment approaches. We use the CPLEX commercial solver for solving our MILP model using the branch-and-bound method, while the method used to solve the LP is primal-dual SIMPLEX. All experiments are carried out on an Intel Xeon processor at $2.3$ GHz with 8GB memory. We consider a 10km x 10km square region and CHs are distributed according to a Matern cluster process \cite{haenggi2012stochastic} with the number of clusters changing between $2-11$. The cluster density mean and cluster radius are $10$ and $1000$m. We use the Reference Point Group Mobility (RPGM) model introduced in \cite{hong1999group}. In this model, GUs in a cluster tend to coordinate their movement and the movement of each CH determines the behavior of the entire group. One example of such mobility is the movement of rescue teams during disaster relief. In our experiments, CHs move according to RWPM and their speed is distributed uniformly according to $U(5, 40) m/s$. We consider a grid with the total number of $100$ points at height $h$ as the potential UAV positions. The ground network flows are generated according to a Bernoulli distribution with the parameter $\lambda = 0.04$ while the traffic demand for each pair is chosen with equal probability among the values $0.2$, $0.4$, and $0.6 Mbps$. Unless stated otherwise, simulation parameters are provided in Table \ref{param-value}. 
\begin{table}[b]
\centering
\caption{Simulation parameters}
\begin{center}
\scalebox{0.9}{
\begin{tabular}{|c|c|}
\hline
\hline
Parameters & Description\\
\hline
 $(a, b ,\eta_{LoS}, \eta_{NLoS})$& $(9.61, 0.16, 1.0, 20.0)$ for urban environment\\
UAVs altitude $h$ & 2000m\\
Carrier frequency $f_c$ & 2GHz\\
Thermal noise power $\sigma^2$ & -90dBm\\
SNR threshold $\gamma_{min}$ & -4 dBm\\
$E_{GU}, E_{UAV}$ & 20dBm, 110dBm\\
$v_{max}$ & 55m/s\\
Snapshot duration $\Delta T$ & 25s\\
\hline
\hline
\end{tabular}}
\end{center}
\label{param-value}
\end{table}
Based on the simulation parameters, the A2G communication range $R_1 = 2214m$ and the corresponding A2A communication range is $R_2 = 3774m$ calculated from equations \eqref{R1} and \eqref{R2}. In the following, the numerical results are provided for two experiments.

\subsubsection{Static ground network}
In this experiment, we consider a fixed network with $9$ clusters and a $10\times10$ UAV location grid. Fig. \eqref{fig:placement_milp}-\eqref{fig:placement_senni} illustrate the UAV placement solution of DM-MILP, DM-LP, DM-Conn strategies. It can be observed that all the clusters are covered by UAVs in all three cases. The number of deployed UAVs in DM-Conn is less than the other two approaches, since D-Conn only ensures the A2G, A2A connectivity and A2G link capacity constraints, not the end-to-end traffic delivery. As a result, the supported traffic of DM-Conn is $67\%$, while both DM-MILP and DM-LP fully support the traffic demands in this example. This experiment highlights the need for joint UAV placement and traffic routing in multi-hop UAV relaying systems.

\subsubsection{Mobile ground network}
In this experiment, the proposed DM-LP is benchmarked against DM-MILP and a static UAV deployment in a dynamic ground netowrk. Fig. \ref{number} depicts the number of relays deployed based on DM-MILP and DM-LP, as an indicator of the deployment cost. The results are averaged over 20 snapshots. There are no hard limits imposed on the maximum number of UAVs, i.e. their number is only constrained by the number of potential UAV positions on the grid ($|\mathcal{U}|$). As a result, traffic demands are fully supported, while the difference in the average number of deployed UAVs is at most $2$ more for the approximation algorithm. With regards to time complexity, as depicted in Fig. \ref{fig:time}, DM-MILP follows an exponential growth, whereas the DM-LP method has an approximately linear time growth with respect to the number of clusters. Under the current evaluation environment, the real time operation of a network comprised of up to 6 clusters can be supported. However, DM-LP's linear time-complexity would guarantee real time support for larger network instances with a more powerful system.

Fig.~\ref{fig:sup} shows the average total supported traffic per snapshot for a ground network of $10$ clusters, following DM-LP, DM-MILP and a static UAV deployment where the UAVs locations are obtained from the solution of DM-MILP for the first snapshot. Fig.~\ref{fig:bp} depicts the profile of the average supported traffic for the same scenarios over snapshots. The results are averaged over $5$ random networks. Overall, the deviation of the DM-LP from the optimal solution is on average $7\%$ and $5\%$ when $N_{max} = 5$ and $N_{max} = 6$ respectively. Note that $6$ UAVs are enough to fully support the traffic demand of the generated network instances as DM-MILP achieves $100\%$ traffic support in this experiment. This demonstrates the ability of the proposed LP-based scheme to generate good solutions, for a limited number of available UAVs. Moreover, a static UAV deployment with even an optimal initial deployment resulted in a maximum of $40\%$ unsupported traffic, highlighting the need for a dynamic UAV deployment solution.

The four figures together shows how the mobility capability of UAVs can be exploited to achieve higher traffic delivery in a mobile ground network and demonstrates that the proposed DM-LP approach trades only a small degree of optimality for fast retrievable solutions.

\section{Related Work}
\label{sec:relatedwork}
Deployment of UAVs has extensively been a topic of research with different objectives, such as the maximization of the downlink (DL) throughput \cite{wu2018joint}, \cite{merwaday2015uav} and DL received signal strength (RSS) \cite{galkin2016deployment}. In contrast, we explore a UAV-assisted communication system considering both uplink (UL) and DL traffic streams. Authors in \cite{al2014optimal}, \cite{mozaffari2016efficient}, \cite{alzenad20173} investigate the usage of UAVs to maximize the covered area with respect to the UAV altitude, antenna gain and minimum received power of users. In \cite{sharma2019random}, the coverage probability of a reference ground user is evaluated for a  3D UAV movement process characterized by the RWPM and uniform mobility models. In \cite{bor2016efficient}, authors propose a UAV-assisted cellular network and maximize the revenue, that is proportional to the number of covered users. However, in order to fully satisfy the QoS requirement of the users in a multi-hop wireless network, the end-to-end traffic delivery should be considered which is more challenging than a coverage problem. Regarding energy efficiency, Mozaffari et al. \cite{mozaffari2016optimal} applied optimal transport theory to minimize total DL transmission power. Optimizing the flight radius and speed to improve energy efficiency is also addressed in \cite{zeng2017energy}.

The required number of aerial UAVs is minimized in \cite{perumal2008aerial}, \cite{kalantari2016number}. Authors in \cite{perumal2008aerial} propose a UAV placement algorithm taking into account the connectivity between UAVs and the clusters demands; however, the constraints on the UAVs mobility and the A2A links capacity are ignored. Authors in \cite{challita2017network} and \cite{li2018uav2} considered multi-hop wireless backhauling in UAV-aided networks. In \cite{challita2017network}, U. Challita and W. Saad seek to form a multi-hop backhaul network in the sky connecting small ground base station through formation of a bidirectional tree structure. Different from \cite{challita2017network}, we consider both A2A and A2G links and jointly optimize the UAV placement and routing. In \cite{li2018uav2}, the authors optimized the UAV placement, power and bandwidth allocation in an UAV-enabled multihop backhaul with fixed number of UAVs. In our case, we minimize the number of deployed UAVs in addition to imposing a constraint on the maximum number of UAVs. In contrast to \cite{multi-relay} which investigates the trajectory design and power allocation strategies for a single fixed ground source-destination, we consider a general mobile ground network consisting of multiple traffic flows which makes the UAV relays trajectory design and traffic routing more challenging and out of the scope of the set-up in \cite{multi-relay}. Authors in \cite{9149409} consider the placement and resource allocation problem of multi UAV relays for a ground network with multiple traffic flows; however they ignore the mobility of the ground nodes and how it affects the UAV locations and other decision variables.
\section{Conclusion}
\label{sec:conclusions}
In this article, we propose a framework for joint UAV placement and route optimization in a multi-hop UAV relaying communications system, taking into account the mobility of the ground nodes, the capacity of A2A and A2G links, UAVs mobility constraints and UAVs propulsion energy consumption. We model the problem as a MILP, and then propose an efficient LP-based approximation algorithm to effectively reduce the time-complexity of our model, achieving a near-optimal solution. The numerical simulations provide insights on the effect the users' mobility and the dynamic relocation of UAVs on the decision making process and the service degradation. Among our future directions are to consider the control/management layer resource allocation problem and investigate the computation offloading and service placement problem together with the resource allocation in a mobile edge computing setup.

\section*{Acknowledgment}
The research of A. Gholami, N. Torkzaban and J.S. Baras, was partially supported by ONR grant N00014-17-12622 and by a grant from the Lockheed Martin Corporation.

\bibliographystyle{splncs04}
\bibliography{bibliography}

\end{document}